\documentclass[preprint]{aastex}
\usepackage{epsfig}
\usepackage{graphicx}
\usepackage[normalem]{ulem}
\usepackage{cancel}
\usepackage{gensymb}
\usepackage{multirow,bigdelim}

\shorttitle{Accretion-Generated Plasma in EX Lupi}
\shortauthors{Teets et al.}

\begin{document}

\title{Detection of a Cool, Accretion Shock-Generated X-ray Plasma in EX Lupi During the 2008 Optical Eruption}

\author{William K. Teets\altaffilmark{1}, David A. Weintraub\altaffilmark{1},  Joel H. Kastner\altaffilmark{4}, Nicolas Grosso\altaffilmark{2},\\
Kenji Hamaguchi\altaffilmark{3}, Michael Richmond\altaffilmark{4} }

\altaffiltext{1}{Department of Physics \& Astronomy, Vanderbilt University, Nashville, TN 37235, USA}
\altaffiltext{2}{Observatoire Astronomique de Strasbourg, Universit\'e de Strasbourg, CNRS, UMR 7550, 11 rue de l'Universit\'e, F-67000 Strasbourg, France}
\altaffiltext{3}{Goddard Space Flight Center, Greenbelt, MD 20771, USA}
\altaffiltext{4}{Rochester Institute of Technology, Rochester, NY 14623-5604, USA}

\begin{abstract}
EX Lupi is the prototype for a class of young, pre-main sequence stars which are observed to undergo irregular, presumably accretion-generated, optical outbursts that result in a several magnitude rise of the optical flux. EX Lupi was observed to optically erupt in 2008 January, triggering Chandra ACIS ToO observations shortly thereafter. We find very strong evidence that most of the X-ray emission in the first few months after the optical outburst is generated by accretion of circumstellar material onto the stellar photosphere.  Specifically, we find a strong correlation between the decreasing optical and X-ray fluxes following the peak of the outburst in the optical, which suggests that these observed declines in both the optical and X-ray fluxes are the result of declining accretion rate. In addition, in our models of the X-ray spectrum, we find strong evidence for a $\sim$0.4 keV plasma component, as expected for accretion shocks on low-mass, pre-main sequence stars.  
From 2008 March through October, this cool plasma component appears to fade as EX Lupi returns to its quiescent level in the optical, consistent with a decrease in the overall emission measure of accretion shock-generated plasma.  The overall small increase of the X-ray flux during the optical outburst of EX Lupi is similar to what was observed in previous X-ray observations of the 2005 optical outburst of the EX Lupi-type star V1118 Ori but contrasts with the large increase of the X-ray flux from the erupting young star V1647 Ori during its 2003 and 2008 optical outbursts.  

\end{abstract}


\section{Introduction}
The classical T Tauri star (cTTs) EX Lupi is the prototype of a class of young, heavily embedded, pre-main sequence (PMS) stars known as ``EXors" that are observed to undergo irregular optical outbursts \citep{her01}, with each outburst continuing for up to a few years.  EX Lupi was first observed to undergo a large optical outburst in 1955, brightening by nearly five magnitudes from a typical quiescent optical magnitude of 13.2 to a peak magnitude of 8.4 before fading back to its quiescent level approximately one year after the onset of the outburst \citep{her77}.  EX Lupi had erupted five times prior since the 1890s; however, these eruptions typically increased the optical brightness of EX Lupi by only one or two magnitudes \citep{mcl46}.  With considerable gaps in the observing data, it is unclear what the peak outburst magnitudes were or how long these outbursts lasted.  It does appear clear, however, that the 1934 eruption lasted much longer (at least six years) than what is typically observed for EX Lupi.  In 1994, almost 40 years after its 1955 outburst, EX Lupi was observed to erupt again, brightening to a peak magnitude of 11.4 in the V band and remaining brighter than magnitude 13.0 for approximately 1.5 years \citep{her01}. Three similar small eruptions occurred during the next eight years with EX Lupi reaching a peak magnitude of 11.3, 11.1, and 10.8 in the V band in 1998 June, 1999 July-July, and 2002 July-August, respectively \citep{her07}.  The most recent eruption began in 2008 January when EX Lupi was observed to brighten to a peak visual magnitude of $\sim$8 at the end of the month \citep{jon08} and remained in an elevated optical-flux state for approximately 8 months.  This latest eruption therefore appears to rival the 1955 ``extreme'' outburst, though the 1955 outburst appears to have lasted a few months longer.

During an EXor outburst, the underlying spectrum is veiled by hot continuum emission and some emission features show an inverse P-Cygni profile \citep{her01}, signifying an infall of circumstellar material.  Consequently, the large-scale variability of EXors is attributed to short-lived increases of the mass accretion rate (e.g., from a quiescent value of $6\times 10^{-9}$ to $2\times 10^{-7}$ $M_\odot$ yr$^{-1}$ for the 2008 optical outburst, according to \citealt{asp10}).  Proposed mechanisms for eruptive young stellar objects (YSOs), such as EXors, include circumstellar disk instabilities that arise because of the gravitational influence of a nearby (unseen) companion \citep{bon92}, thermal instabilities in the circumstellar disk \citep{bel91,cla89}, and magnetohydrodynamic instabilities within the disk \citep{arm01,dan12}.  
EXors are similar in behavior to another class of young, erupting stars known as ``FUors.''  These stars are also thought to erupt due to sudden, massive accretion events.  The main differences between the two classes are that FUors have higher accretion rates (roughly $10^{-4}$ $M_\odot$ yr$^{-1}$) during outbursts, resulting in the visible flux levels being elevated by several more magnitudes than what is found in EXor outbursts, and that the durations of the FUor outbursts are typically decades whereas EXor outbursts are, at most, a few years in duration \citep{har96}.

PMS stars are known to produce a significant amount of flux as X-rays (L$_{X}$/L$_{bol}$ $\sim$ 10$^{-4}$--10$^{-3}$, compared with L$_{X}$/L$_{bol}$ $\sim$ 10$^{-7}$--10$^{-6}$ for the Sun), although the  origin of the X-ray flux is not well understood.  The hard (E $\gtrsim$ 1.0 keV) X-ray flux is generally attributed to coronal activity, as in the analysis by \citet{pre05} of observations of a large sample of young stars in Orion in the Chandra Orion Ultradeep Project (COUP).  Others have found that accretion of material onto the star could be another source of soft (E $\lesssim$ 1.0 keV) X-rays.  Circumstellar material that is accreted onto the stellar photosphere at near free-fall velocity is capable of generating soft X-rays as it is heated to a temperature as high as a few million degrees Kelvin (kT$_{X}$ $\sim$ 0.04 to 0.4 keV) \citep{cal98,sac1}.  This emission can be now identified via X-ray spectroscopy (see, e.g., review by \citet{gud09}).  An excellent example of an object in which this process appears to be occurring is the nearby, relatively evolved (age $\sim$ 10 Myr) cTTs TW Hydrae \citep{kas02,brick10,dup12}.  \citet{brick10} modeled the accretion onto a stellar photosphere and found two regions where soft X-ray emission could arise:  an inner shock front, where material first impacts the photosphere, and a outer post-shock region that is heated by energy transfer from the shock region to coronal material at temperatures above 1MK.  To further complicate the issue, some accreting stars may have their X-ray production, specifically the softer X-rays, quenched during accretion episodes \citep{gre07}.  Quenching could occur if the accretion column absorbs the accretion shock-generated X-rays.

The PMS star V1647 Ori exhibited different behavior when it was observed by CXO and XMM-Newton during its optical/NIR outbursts from 2002--2006  and again from 2008--2009 \citep{kas04,kas06,gro05,muz05,gro06,ham10,tee11}, when the mass accretion rate varied from a quiescent value of $6\times 10^{-7}$ to $2\times 10^{-5}$ $M_{\odot}$ yr$^{-1}$.  During the optical outbursts of V1647 Ori in these two epochs, the X-ray flux correspondingly increased by up to two orders of magnitude; however, the increases in X-ray flux were primarily detected in the harder (2.8--8.0 keV) energy range.  The spectral characteristics of V1647 Ori during both outbursts were remarkably similar and best modeled as plasma at temperatures on the order of 2--6 keV, too high to be produced by an accretion shock but in the range expected from magnetic reconnection events in the accretion streams \citep{tee11}.  The high density of the gas around V1647 Ori causes the local gas column to be nearly opaque to lower-energy X-ray photons, making the direct detection of X-ray flux from shocks at accretion hotspots difficult; 
however, \citet{ham12} suggest that the observed modulation of the hard X-ray flux from V1647 Ori is the signature of accretion hotspots rotating in and out of the field of view.

EX Lupi was observed in the UV and X-ray regimes with XMM-Newton during a 78 ks observation beginning on 2008 August 10.  \citet{gro10} found the X-ray spectrum of EX Lupi to be best modeled as a two-component plasma with one component at a temperature of kT$_{X}$ $\sim$ 0.5 keV suffering very low extinction from a hydrogen column density of 3.6 $\times$10$^{20}$ cm$^{-2}$ and the other, which dominated the intrinsic X-ray emission, at a temperature of kT$_{X}$ $\sim$ 4.6 keV suffering very high extinction from a hydrogen column density of 2.7 $\times$10$^{22}$ cm$^{-2}$.  The cooler component was determined to likely be associated with X-ray emission from accretion shocks due to its low X-ray temperature, and the UV activity observed was found by \citet{gro10} to be typical of accretion events and  dominated by emission from accretion hotspots covering about one percent of the stellar surface.  

In this paper, we present three periods of X-ray observations of EX Lupi obtained with the Chandra X-ray Observatory (CXO). The first X-ray dataset was obtained approximately two months after the peak of optical outburst, the second was collected three months later, and the third about seven months after the first observation.  These three observation epochs, along with the XMM-Newton observation in 2008 August, allow us to follow the X-ray evolution of this object as it faded back toward quiescence at optical wavelengths.  In \S2, we describe the observations and data reduction.  In \S3, we discuss the results and their implications in the context of accretion shock-generated X-ray production.  

\section{Observations \& Data Reduction of X-ray Data}

A 20.1 ks exposure observation of EX Lupi was triggered on 2008 March 25 (CXO Cycle 9, anticipated Target of Opportunity; PI:  D. Weintraub, ObsID 8923) with CXO after EX Lupi was observed to be in optical outburst in 2008 January \citep{jon08}.  Subsequent program observations were initiated on 2008 June 16 (ObsID 8924, 20.1 ks), 2008 October 6 (ObsIDs 8925 and 10789 with durations of 10.6 and 15.4 ks,  respectively), and 2008 October 9 (ObsID 10791, 4.1 ks).  The 2008 October observation was split into three separate exposures due to CXO scheduling constraints.  ObsIDs 8925 and 10789 were both obtained on 2008 October 6 while ObsID 10791 was obtained on 2008 October 9.  Together, these five pointings yield an observing sequence, spanning approximately seven months, that follow the X-ray evolution of EX Lupi from two months after the start until after the conclusion of the optical outburst.  During this same seven-month period (specifically in 2008 August), a 78 ks observation of EX Lupi was also obtained by XMM-Newton (AO-7, anticipated Target of Opportunity; PI: N. Grosso, \citealt{gro10}).  For all Chandra observations, the Advanced CCD Imaging Spectrometer Imaging (ACIS-I) array was used in faint telemetry mode with EX Lupi at the aimpoint of the I3 CCD.

\begin{figure}

\centering
\begin{tabular}{cc}
\includegraphics[bb=30 10 360 350, scale=.7]{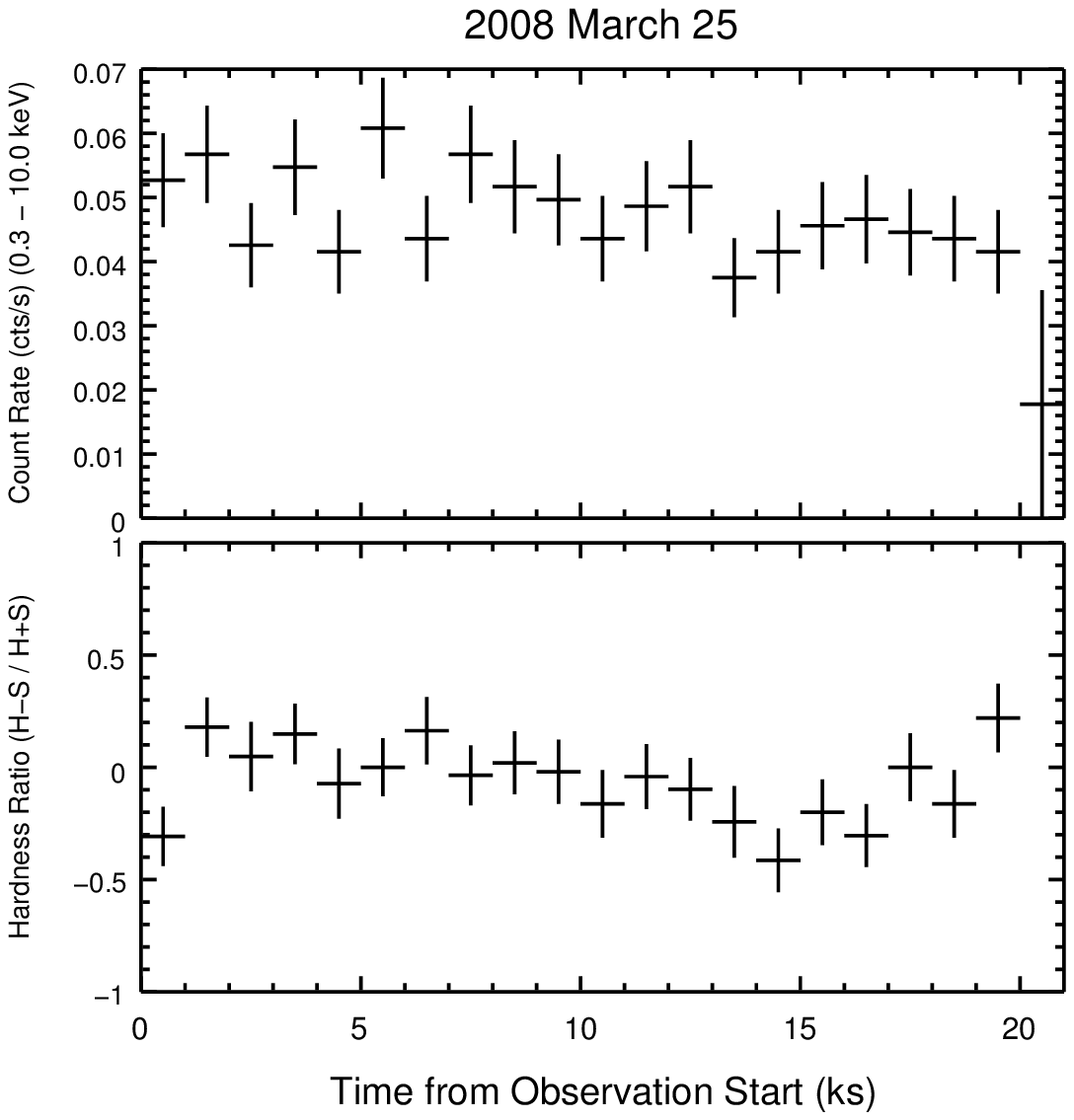} & \includegraphics[bb=30 10 360 350, scale=.7]{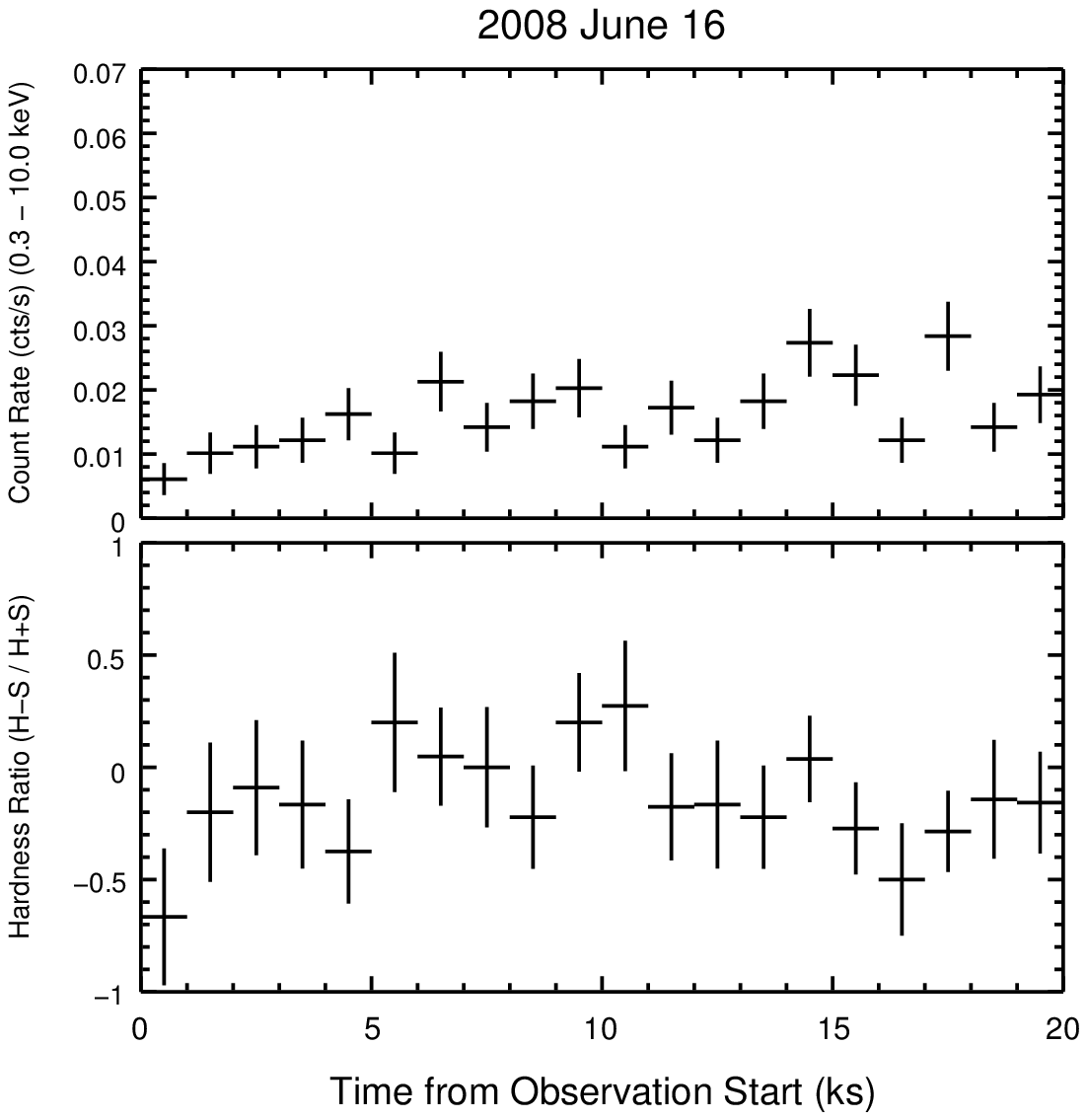}
\end{tabular}
\begin{tabular}{c}
\includegraphics[bb=30 10 760 350, scale=.7]{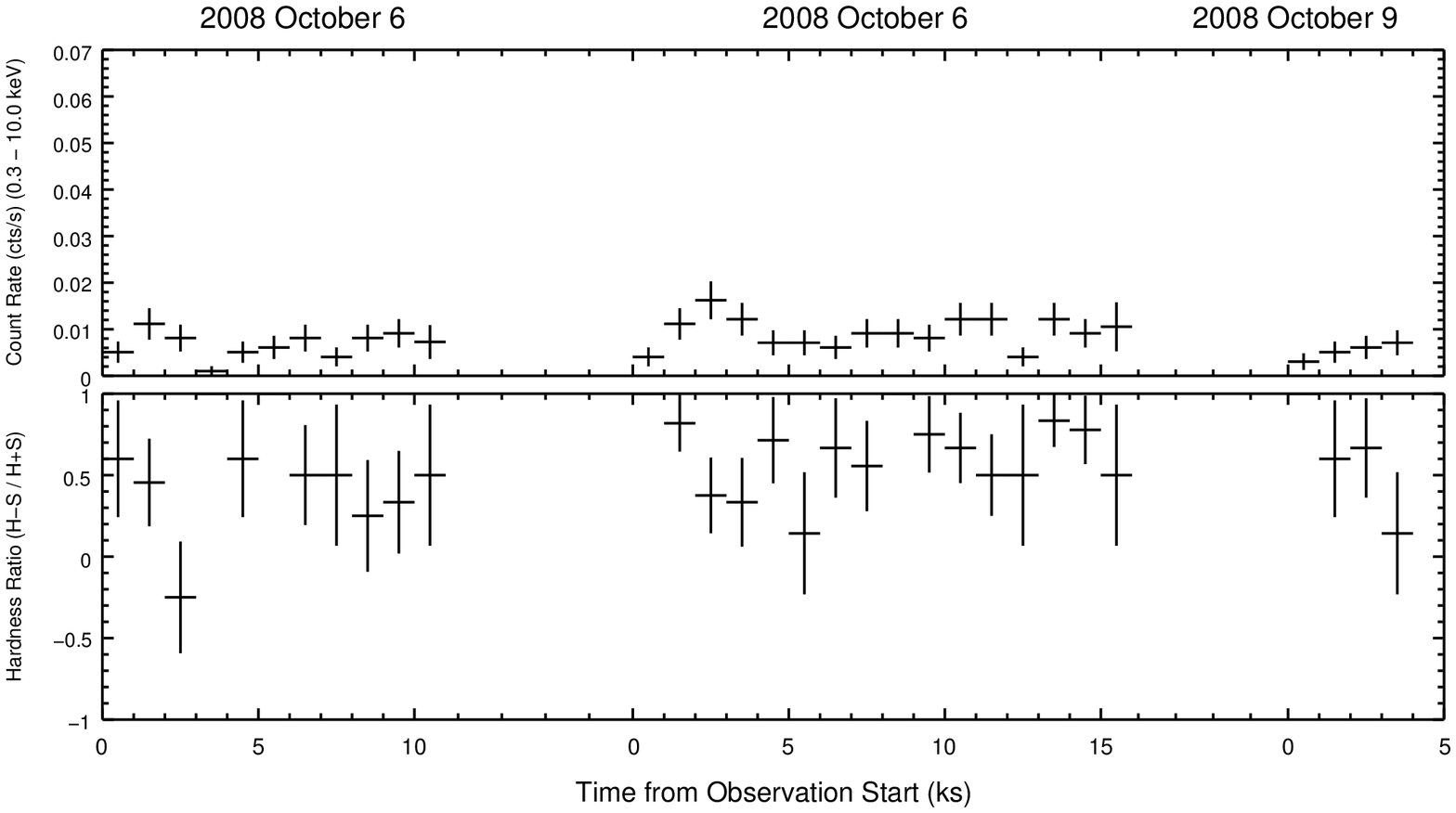} 
\end{tabular}
\caption{Background-subtracted X-ray light curves (top panes of each panel) and hardness ratio curves (bottom panes of each panel) for the 2008 CXO observations of EX Lupi.  Time bins are 1 ks each, and the soft and hard energy bands are 0.3--1.2 keV and 1.2--10 keV, respectively. Plotted uncertainties in hardness ratios and count rates are 1$\sigma$.  The light curves and hardness ratio curves for the three 2008 October observations have been concatenated in the bottom panel with arbitrary time gaps between observations.\label{lchr}}
\end{figure}

CIAO v4.1 and CALDB v4.1.0--4.1.4 were used to reduce the data and extract pulse-invariant (PI) spectra.  Observation details are given in Table~\ref{tab-1}.  CXO/ACIS has a calibrated energy range of 0.3--10 keV, so the observation event files were first filtered to only include events with nominal energies that fell within this range.  Source spectra were then extracted from 2.5$\arcsec$ radius regions (making sure the aperture size was appropriate to encompass $\gtrsim$90\% of the photons) while background spectra were extracted from regions near but beyond 2.5$\arcsec$ from the target, on the same CCD (I3), using 20$\arcsec$ outer radius extraction apertures. The light curves for all of these CXO observations are shown in Figure \ref{lchr}.  The hardness ratios do not change in any significant way, and no flaring events appear in these data.  In October, the X-ray count rate appears to be at roughly the same level for all three observations (Figure \ref{lchr}, bottom panel). In addition, the median photon energy (not shown) is statistically indistinguishable in the October data sets.  Finally, the October observations were obtained close enough in time to one another that the detector characteristics/responses of ACIS should not have changed in any significant way.  Therefore, we used CIAO to extract and combine the spectra of the three October observations and present the composite X-ray spectrum characteristics, along with those of the 2008 March and June spectra, in Table \ref{tab-1}.  The spectra were grouped into energy bins with a minimum of five counts per bin prior to spectral modeling.  The count rates were high enough and durations long enough for each of the observations that this bin size yielded PI spectra with good statistics.

\section{Results from X-ray Observations}


\subsection{General Spectral Characteristics Indicating Accretion}\label{spec_char}

Our analysis indicates that X-ray spectrum of EX Lupi consisted of a bright source of relatively soft X-rays immediately after the outburst but had changed to a faint source of harder X-rays by the end of the outburst.  Table \ref{tab-1} gives the mean count rates, median photon energies, and hardness ratios of the spectra of EX Lupi for the 2008 CXO observations, and Figure \ref{crehr} illustrates the temporal changes in these properties of the X-ray spectra and the correlation between the decline in the V band and the decrease in X-ray flux.  Immediately following the onset of the optical outburst, the X-ray count rate is high, the median photon energy is soft, and the hardness ratio\footnote[1]{Hardness ratio = (H--S/H+S), where S is defined as the soft X-ray band (0.3--1.2 keV) and H is the hard X-ray band (1.2--10.0 keV).} is modestly negative.  Approximately three months later, the X-ray count rate had decreased by a factor of $\sim$3, and the median photon energy and hardness ratio were roughly unchanged.  In 2008 October, approximately six months after the first X-ray observation of EX Lupi and roughly eight months after the onset of the optical outburst, the X-ray count rate had declined to roughly one-sixth of its 2008 March 25 value, the median photon energy had increased by a factor of $\sim$1.5, and the hardness ratio had become significantly positive.


\subsection{Spectral Modeling}\label{modeling}

In modeling the EX Lupi spectra, we employed XSPEC v12.4.  First, we computed models with a thin, single-temperature plasma simulated with an APEC component \citep{smi01} subject to absorption by an intervening column of gas (WABS component - see \citealt{mor83}).  
Chemical abundances were set to values found in the XEST survey \citep{gud07}, consistent with the approach in \cite{gro10}. Initial model parameter values were chosen to lie within the parameter range found by \citet{gro10}.  Next, we fit the data to a two-component model with the second plasma component subject to the same hydrogen column density.  Finally, we again reset the two-component plasma model to the same starting parameters and incorporated a second intervening hydrogen column.  In all three models, the hydrogen column densities and plasma temperatures were allowed to vary while the chemical abundances remained fixed. 

\subsubsection{2008 March}\label{march2008}
\begin{figure}
\centering
\includegraphics[scale=.70, angle=270]{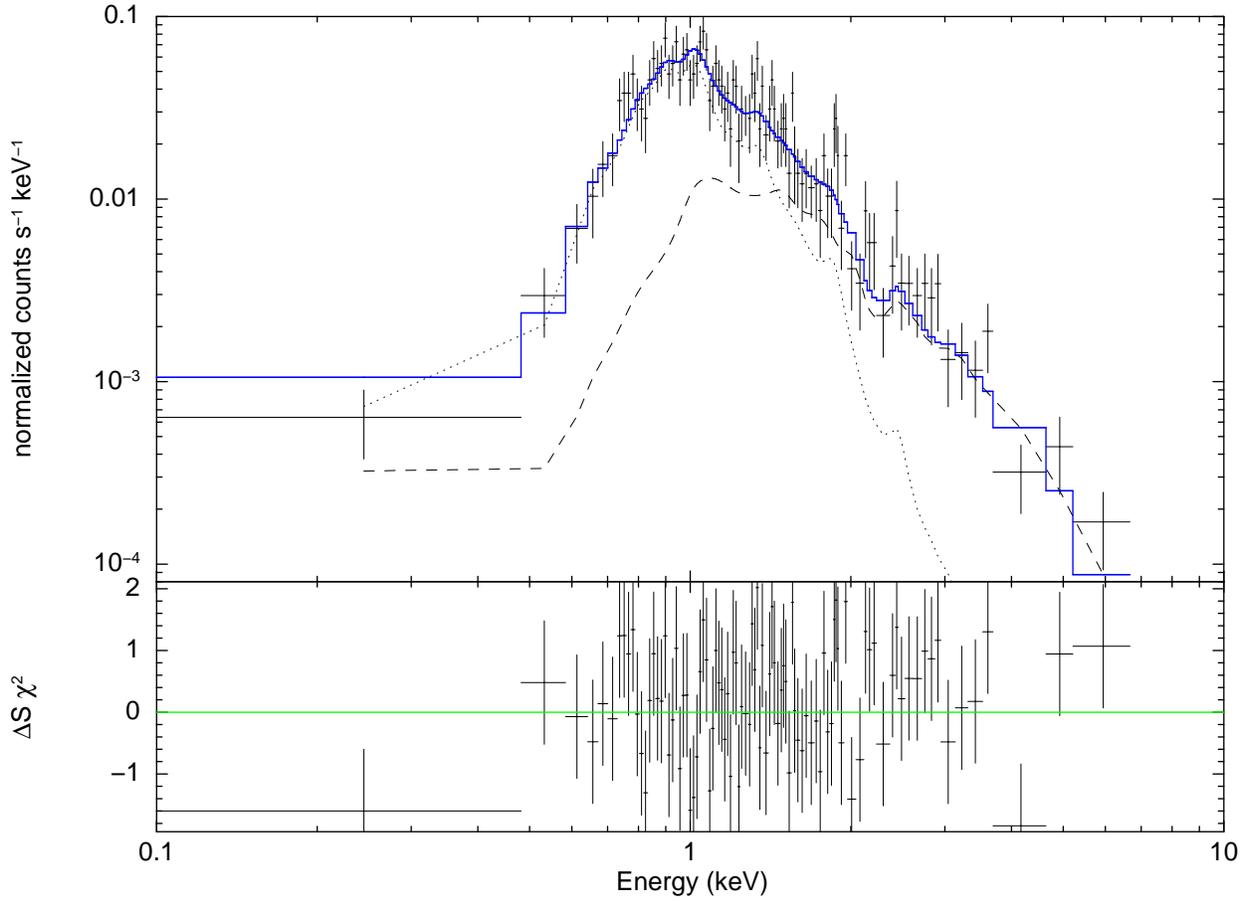}\\  
\caption{The best-fit two-component XSPEC model of the 2008 March EX Lupi spectrum (data binned to five-count-minimum bins) is displayed (blue line) in the top panel.  Residuals to the fit are given in the bottom panel.  The horizontal lines for each of the data points represent the bin widths, and the vertical lines represent the one-sigma error bars.  The dotted and dashed lines represent the absorbed contributions from the lower-temperature and higher-temperature plasma components, respectively.}
\label{8923modelcomp}
\end{figure}

The spectrum of ObsID 8923 (Fig. \ref{8923modelcomp}) was obtained approximately two months after EX Lupi was first observed to be in an eruptive state.  We find that this spectrum is fit best (F-test probability\footnote[2]{F-test probability provides an assessment of the improvement in using one model versus another and is calculated from the chi-squared and degrees-of-freedom values of the two model fits using the XSPEC \emph{ftest} command.  An F-test probability much less than unity (usually $\lesssim$0.05) suggests that it is very reasonable to add the additional model component. } = 4.5 $\times$10$^{-6}$) with the two-component plasma (``2T'') model (Table \ref{tab-3}) with temperatures of 0.4 keV and 1.7 keV suffering extinction from a single intervening absorbing column of N$_{H}$=0.4 $\times$10$^{22}$ cm$^{-2}$.  The lower-temperature component contributes roughly four times more X-ray flux (as derived from absorption-corrected models) than the higher-temperature component and accounts for most of the emission up to energies of $\sim$1.5 keV.  The addition of another parameter describing the absorption toward the second (hotter) plasma component does not improve the fit (F-test probability = 0.92) for this particular observation, as the value of the second hydrogen column density converges to the same value as the first.  A single-component plasma (``1T'') model can account for the X-ray emission at energies below 2 keV but is unable to account for most of the X-ray emission from $\sim$2 to 6 keV.

\subsubsection{2008 June}\label{june2008}

\begin{figure}
\centering
\includegraphics[scale=0.7, angle=270]{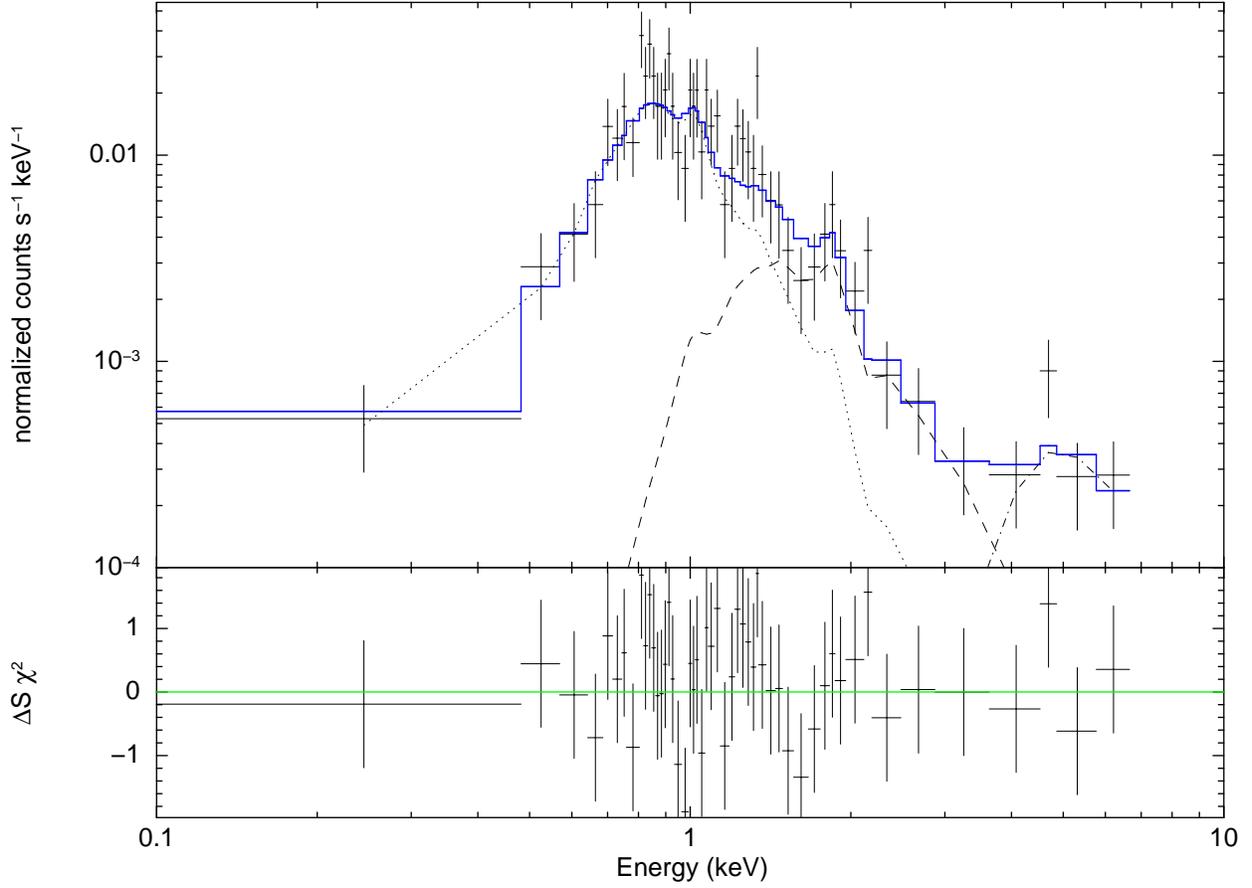}\ 
\caption{The best-fit three-component XSPEC model of the 2008 June EX Lupi spectrum (data binned to five-count-minimum bins) is displayed (blue line) in the top panel.  Residuals to the fit are given in the bottom panel.  
The remaining lines represent the absorbed contributions from the 0.5 keV (dotted line), 0.7 keV (dashed line), and $\sim$4 keV (dash-dotted) plasma components.\label{8924modelcomp}}

\end{figure}

The spectrum of ObsID 8924 (Fig. \ref{8924modelcomp}) is best fit with a three-component plasma, including a heavily absorbed, high temperature component that is not evident in the March (\S3.2.1), August \citep{gro10}, or October (\S3.2.3) spectra.  When compared to the other CXO and XMM-Newton observations of the X-ray spectrum of EX Lupi, the June spectrum appears to have ``excess'' X-ray flux at energies above $\sim$4 keV (see \S3.3).  However, closer inspection reveals that the emission above $\sim$4 keV can be attributed to a hot X-ray plasma component that is present during all 2008 observations of EX Lupi (see \S3.3.2).

A 1T model was unable to account for emission above $\sim$2 keV while a 2T model resulted in a better fit of the spectrum above $\sim$2 keV but a poor fit to the spectrum at energies greater than $\sim$4 keV.  We note that this 2T model converged to a fit with a very low hydrogen column density (0.05 $\times$10$^{22}$ cm$^{-2}$) in comparison to the hydrogen column density that provided the best fit to the 2008 March spectrum, though this low column density is very similar to one of the hydrogen column densities found by \citet{gro10} for the 2008 August X-ray spectrum.  The temperature of the hotter plasma component in this model (0.5 keV and 11.6 keV) is not well constrained, and we find that the emission above $\sim$4 keV is not well accounted for.  The addition of a second hydrogen column density resulted in the fit of the hotter plasma component diverging to an unrealistic temperature.  
 We conclude that this two-component model is not adequate for modeling this spectrum.  

We then fit the spectrum with a three-component plasma model with each plasma component subject to a separate hydrogen column density (Table \ref{tab-3}).  In order to have some control over the fitting algorithm, we first modeled the spectrum up to the $\sim$4.5 keV energy range (the portion of this X-ray spectrum that was visually similar to the 2008 March spectrum) with a two-temperature/two-hydrogen column density model, which yielded a good fit to this portion of the spectrum.  We then fixed those parameter values, reset the modeled energy range to include all of the energy spectrum, and added the third hydrogen column density and plasma components to the model.  
We held the chemical abundances fixed but allowed all other parameters to vary and reran the model, allowing the parameters to adjust to find the best fit.  The addition of the third plasma component improved the fit for the entire spectrum (F-test probabilities, when going from a 1HCD/2T model and a 2HCD/2T model to a 3HCD/3T model, are 0.08 and 0.05, respectively, which suggests that the 3HCD/3T model is an improvement over both the 1HCD/2T and the 2HCD/2T models).  The $\sim$4 keV plasma component does not have a well-constrained temperature or hydrogen column density, but we find that the higher-energy portion of the 2008 June spectrum is not adequately fit without the additional plasma component \emph{and} hydrogen column density.  Therefore, we conclude that our 2HCD/2T model fits the 2008 June 16 spectrum well up to energies of $\sim$4 keV; however, the models show that this spectrum does require a third, heavily-absorbed plasma component in order to characterize it more completely.


As shown in Figure \ref{8924modelcomp}, the prominent emission above $\sim$4.5 keV is contributed by the most heavily-absorbed plasma component with a plasma temperature of $\sim$4 keV.  The additional hydrogen column density, which is over an order of magnitude greater than the second-highest column density of this observation, is required to quench the lower-energy portion of the spectrum of the $\sim$4 keV plasma such that only the higher-energy portion is able to contribute to the 2008 June spectrum.  We also find that the light curve of ObsID 8924 (Figure \ref{lchr}, top right panel) did not indicate any large-scale variability during the 20 ks observing period, and the spectrum hardness ratio throughout this observation remained at a constant level.  Therefore, it does not appear that there were any flaring events that might have contributed to the appearance of the third plasma component during only a short portion of the observation.  One interpretation of this spectral behavior is that the accretion rate had decreased enough by 2008 June 16 that we were able to detect the X-ray signature of the corona itself; this interpretation is discussed later in Section \ref{post-outburst_plasma}.

\subsubsection{2008 October}\label{october2008}

\begin{figure}
\centering
\includegraphics[scale=0.7, angle=270]{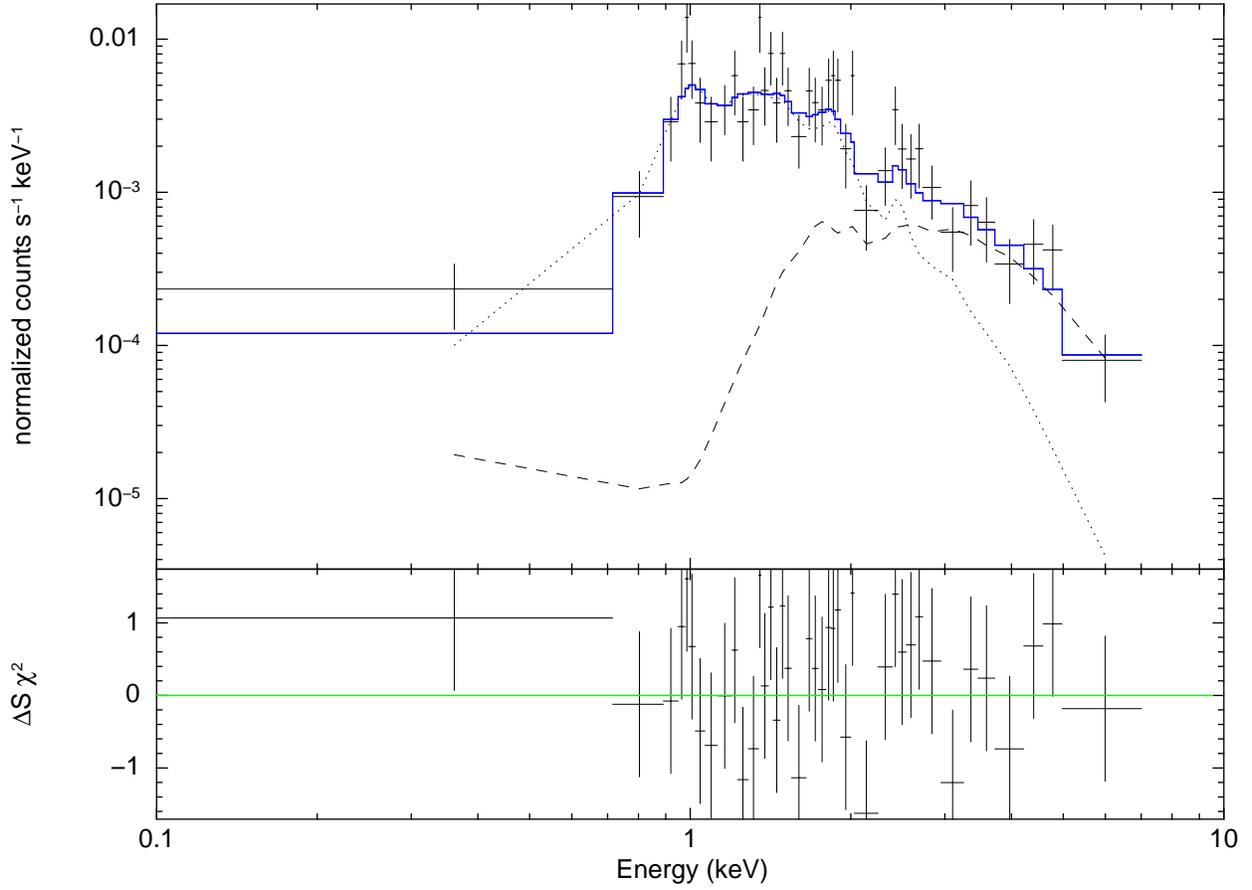}\
\caption{The best-fit XSPEC model of the 2008 October EX Lupi composite spectrum data (binned to five-count-minimum bins) is displayed (blue line) in the top panel.  Residuals to the fit are given in the bottom panel.  The dotted and dashed lines represent the absorbed contributions from the lower-temperature and higher-temperature plasma components, respectively.    
\label{octoberspectra}}
\end{figure}

 We find that the best-fit model for the 2008 October combined spectrum of EX Lupi (Figure \ref{octoberspectra}) is a two-component plasma with the two components subject to absorption by gas with different hydrogen column densities (Table \ref{tab-3}).  The F-test probabilities of using a 2HCD/2T model instead of 1HCD/1T and 1HCD/2T models were 0.08 and 0.05, respectively.  The best-fit model includes both a low-temperature plasma with a modest column density and a high-temperature plasma with a larger column density.  The emission measures and unabsorbed plasma X-ray fluxes are comparable for the two components; however, the best-fit values of these parameters are significantly higher than those of the cooler plasma component during the 2008 August 10 observation and roughly twice as high as those of the hotter plasma component during that exposure.

\subsection{The Temporal Evolution of the Post-Outburst X-ray Plasma}\label{post-outburst_plasma}    
\subsubsection{Change in Plasma Temperature and Emission Measure}\label{temp_em}

From Figure \ref{crehr}, we see that EX Lupi began to gradually fade back toward quiescent optical levels immediately after the onset of the optical outburst in 2008 January. Comparing this trend to what was observed in the X-ray spectral characteristics as EX Lupi returned to optical quiescence (Tables \ref{tab-1} and \ref{tab-3}), we see correlations that strongly suggest that the X-ray evolution and optical evolution of EX Lupi are linked.  

The 2008 March X-ray spectrum of EX Lupi is best modeled by two plasmas, one with a temperature of kT$_{X}$=0.4 keV and another with a higher temperature of kT$_{X}$=1.7 keV.  The lower-temperature component has an absorption-corrected flux  and emission measure that are roughly four times higher than the flux and emission measure associated with the higher-temperature component, and its temperature is characteristic of plasma heated in shocks due to accretion of circumstellar material onto the stellar photosphere.  

Roughly three months later, EX Lupi had declined modestly ($\sim$0.05 magnitudes) in optical flux while the X-ray count-rate had dropped to roughly one-third of the 2008 March level.  In 2008 June, the X-ray emitting plasma is best modeled by three different components, two of which were of similar low temperature but subject to absorption by different hydrogen column densities.  Given their derived 90\% confidence intervals, the lowest-temperature plasma (kT$_{X}$=0.5 keV) could have been generated via an accretion hotspot, but the second, low-temperature plasma (kT$_{X}$=0.7 keV) is too hot to have been generated by an accretion hotspot. This plasma temperature is consistent with the cooler plasma component found by \citet{pre05} in their X-ray spectral fits of many of the COUP sources.  They found that for nearly all stars in their ``optical sample'' the cooler plasma components of their two-component plasma models had temperatures of $\approx$8--10 MK (kT$_{X}$ $\approx$ 0.7--0.9 keV). \citet{brick10} also found evidence of a similar plasma component in TW Hya from the presence and ratios of certain emission lines; they interpreted that this component was possibly due to ``accretion-fed coronal loops'' with temperatures around 10 MK.    The third plasma component had an X-ray temperature of $\sim$4 keV, which is also too high to be generated at an accretion hotspot; however, it is possible that this hotter plasma could be due to magnetic reconnection events in the accretion stream, as we proposed for V1647 Ori \citep{tee11}.  

We find that the coolest plasma of the 2008 June 16 observation had an emission measure that was approximately one-fourth that of the other, hotter plasmas. This is in stark contrast to the results of the 2008 March 25 spectral model, which showed the opposite  --- a cooler plasma with roughly four times the emission measure and absorption-corrected X-ray flux as that of the hotter plasma.  This suggests that the amount of the cooler, accretion shock-generated X-ray plasma had decreased significantly over a span of three months; however, since there had only been a very modest decrease in the optical flux of EX Lupi by 2008 June 16, accretion apparently was still occurring during this CXO observation.  Though there had been a sharp decline in softer X-ray flux (by a factor of $\sim$3) since 2008 March 25, the overall level of X-ray flux remained quite high (as shown by the absorption-corrected flux levels), including a significant contribution from the heavily-absorbed third plasma component.

From 2008 June to August, the plasma temperature appears to have remained largely unchanged, although the emission measure dropped dramatically.  \citet{gro10} found evidence for two X-ray plasmas in the 2008 August spectrum, one of which (0.4 keV) is nearly identical to the 0.4 keV plasma seen in March and the 0.5 keV plasma seen in June, which we have also identified as likely generated at an accretion footprint.  The emission measure of this low-temperature plasma, which had dropped by two-thirds from March to June, by August had dropped to about one percent of the original (March) value.  The accretion phase for EX Lupi, as seen in X-rays, apparently was nearing its end.  This conclusion is supported by the optical measurements, in which the V band magnitude had dropped at least 1.5 magnitudes from March to August and was in the middle of a precipitous descent back to its quiescent magnitude, which it would reach in only ten more days (Figure \ref{crehr}).

In August, during the low-level period observed with XMM-Newton, the hotter plasma component had a temperature and emission measure (within the 90\% confidence intervals) very similar to the hotter plasma component that was observed in the 2008 March 25 spectrum of EX Lupi.  Given that its temperature is too high to be generated by an accretion shock and that the optical flux had returned to its quiescent level, the hotter plasma component is most likely associated with coronal activity. Moreover, in August, this hotter plasma component exhibited flaring activity \citep{gro10}.

The 2008 October spectrum of EX Lupi is best modeled with two plasma components; however, even taking into consideration the 90\% confidence intervals of the modeled plasma temperatures, we find that the two plasmas are too hot to be generated by an accretion hotspot.  Thus, the hotspot plasma signature detected in the spectrum of EX Lupi in the months prior had declined to levels below the detection threshold, supporting the conclusion that once the optical eruption had ended, the X-ray flux had returned to levels typical for EX Lupi in a low accretion state.  The higher-temperature plasma component and its associated column density are similar to the high temperature plasma component seen in both June and August.  The total emission measure in October is only about 20\% that of March.  These results suggest that the several-keV plasma that is associated with a total emission measure of about $\sim$8 $\times$ 10$^{23}$ cm$^{-3}$ is the recovered X-ray signature of the quiescent corona of EX Lupi and that its spectrum likely includes a ``quiescent mass accretion rate'' component that is too low to produce a large observable excess in the CXO X-ray spectrum.

\subsubsection{A Link Between the Accretion Outburst and the Observed Plasma Characteristics}

It appears that during an optical eruption, the X-ray spectrum of EX Lupi is dominated by a cool plasma with a temperature characteristic of an accretion hotspot.  As the young star fades in the optical toward a quiescent level, the X-ray spectrum also changes such that it then becomes dominated by accretion-fed plasmas.  Finally, once the star is at an optically-quiet level, the X-ray spectrum has evolved such that it is dominated by X-ray-emitting plasmas that are characteristic of coronal activity.  

From 2008 March to October, EX Lupi went from an optically-eruptive state to an optically-quiescent state as well as from an X-ray-eruptive state to an X-ray-quiescent state.  Near the onset of the accretion-fed optical eruption, the X-ray spectrum was dominated by a low-temperature plasma characteristic of an accretion hotspot and also included a hotter plasma component that is characteristic of coronal activity.  By 2008 June, EX Lupi had dimmed slightly in the optical, the cool plasma, which had dominated the 2008 March X-ray spectrum, was still present but had drastically decreased in emission measure, and the X-ray spectrum was dominated by plasmas too hot to be generated in accretion hotspots.  Roughly two months later, EX Lupi had dimmed in the optical to a near-quiescent level. The cooler plasma component, which still had a temperature characteristic of accretion hotspots, had decreased further in emission measure while the hotter plasma component, which had a temperature typical of coronal activity, remained relatively unchanged in emission measure.  Finally, by 2008 October, EX Lupi had returned to an optically-quiet state, and the only detectable X-ray plasmas had temperatures that were characteristic of coronal activity --- the signature of the plasma from the accretion footprints was no longer observable.  

\begin{figure}
\centering
\includegraphics[scale=.7, angle=270]{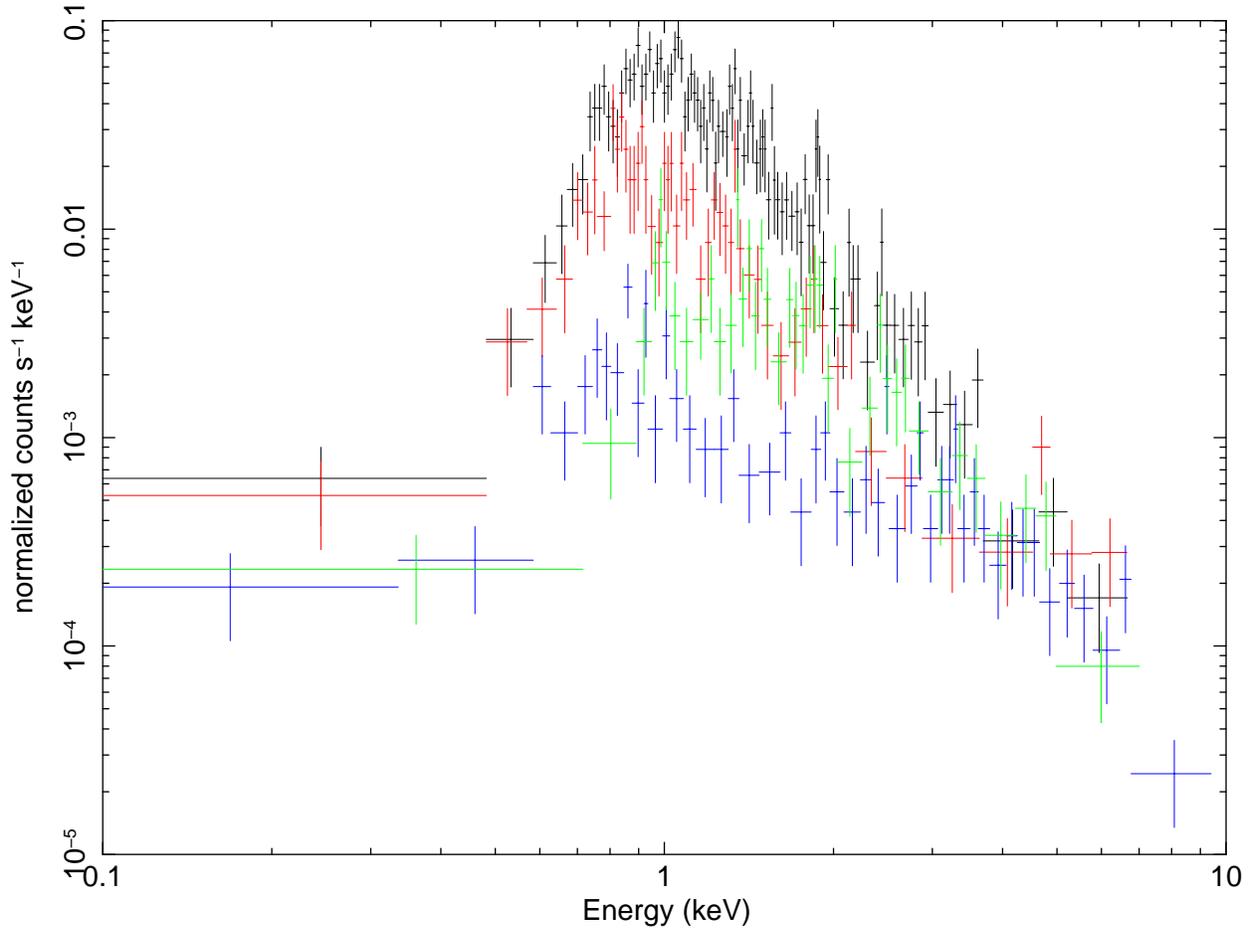}\\
\caption{Overlay of the three 2008 EX Lupi CXO spectra (March - black, June - red, and October - green) and the 2008 August XMM-Newton spectral model of the low-level period (blue) from \citet{gro10} convolved with a CXO response.  While the softer portion of the spectrum of EX Lupi decreases over time as accretion diminishes, the 4--10 keV range appears mostly unchanged over the same time interval (i.e., the 2008 June 16 X-ray spectrum does not seem to have an excess above $\sim$4 keV).  Thus, it appears that we are detecting the X-ray emission from the active corona of EX Lupi in this spectral range.
\label{overlay}}
\end{figure}

Figure \ref{overlay} graphically presents the three CXO spectra of EX Lupi overlaid with a simulated spectrum of the 2008 August XMM-Newton (low-level period in Table 1 of \citealt{gro10}) after being convolved with a CXO response.  We see that over time the softer portion ($\lesssim$4 keV) of the X-ray spectrum of EX Lupi diminishes as the accretion rate decreases; however, the harder portion remains at roughly the same level.  Thus, while it appears that there is an X-ray ``excess'' at energies above ($\gtrsim$4 keV) in the 2008 June X-ray spectrum of EX Lupi, it appears that the excess is actually the X-ray signature of the active coronal plasma, which is otherwise present at a nearly constant level during all observing epochs.

The fits of the 2008 June, August, and October observations of EX Lupi all improve with spectral models that have more than one hydrogen column density component.  It is likely the case that the contribution of the lower-temperature plasma during the 2008 March 25 observation overwhelmed the signal at the lower end of the energy spectrum for the second, higher-temperature plasma and that, coupled with the level of noise in the spectrum during the observation, the derived single X-ray-absorbing column was adequate to account for the absorption suffered by both plasmas.  By 2008 June, however, the X-ray flux from the dominant, lower-temperature plasma had decreased enough that the lower-energy end of the spectrum of the higher-temperature plasma became noticeable and our modeling was sensitive enough so as to require the presence of a second parameter in order to account for the intervening absorbing column toward the hotter component. 

\subsubsection{EX Lupi Compared to Other Young Stellar Objects in Optical/NIR Outburst}

The evidence suggests that the X-ray flux of EX Lupi increases dramatically as the mass accretion rate is enhanced during the optical outburst.  This conclusion is evident when we compare the behavior of EX Lupi to that of two other protostars, V1647 Ori and V1118 Ori,  during X-ray flaring episodes.

The observed trends in EX Lupi (cooler X-ray emitting plasma during outburst than during quiescence; X-ray luminosity increased by factor of 4--5) are in sharp contrast to what was observed during multiple X-ray and optical/near-infrared observations of the erupting pre-main sequence star, V1647 Ori.  During both the 2003 and 2008 optical eruptions, observations of V1647 Ori revealed that the X-ray plasma temperature was on the order of $\sim$4 keV while the X-ray plasma temperature was significantly cooler during optical quiescence ($\sim$1 keV) \citep{tee11}.   The outburst mass accretion rate of V1647 Ori was also about 1,000 times larger than that of EX Lupi during outburst. This high level of accretion should have generated a large number of accretion-fed loops, which are reservoirs of hotter plasma than what would be found in accretion footprints \citep{brick10} and are locations in which magnetic reconnection events would occur.  In addition, the optical flux of V1647 Ori varied by a factor of 60--100 between quiescent and eruptive phases while the X-ray flux varied more by a factor of 150--500, suggesting that although X-rays are the result of accretion in both cases, the observed X-ray-generating sources were different for these two eruptive pre-main sequence stars.  

On the other hand, the behavior of EX Lupi during the 2008 outburst is very similar to what was observed in the X-ray and optical/near-infrared behavior during the 2005 optical outburst of V1118 Ori.  For the duration of the 2005 optical outburst, the optical and near-infrared fluxes of V1118 Ori varied by a factor of 2--10 while the X-ray flux was found to vary only by a factor of 2.  The estimate of the mass accretion rate of EX Lupi during the 2008 optical outburst is very similar to the value of (4--7)$\times10^{-7}$ M$_\odot$ yr$^{-1}$ derived by \citet{lor09} for V1118 Ori (on 2005 September 10, i.e., during the outburst) from the luminosities of emission line Pa$\beta$ and Br$\gamma$.  In addition, the X-ray-emitting plasma of V1118 Ori was shown to be cooler during the outburst than during optical quiescence, suggesting that the X-ray plasma temperature change was likely due to enhanced accretion onto the star \citep{aud05,aud10}.  During the 2008 outburst of EX Lupi, we observed that the optical flux varied by a factor of 50--100 while the X-ray flux changed by a factor of 4--5.  As EX Lupi returned to optical quiescence, the X-ray plasma temperature generally increased.  Once EX Lupi had returned to optical quiescence, the X-ray plasma temperature was similar to that observed for V1118 during optical quiescence.  Given the very similar X-ray behaviors of EX Lupi and V1118 Ori during optical outburst and quiescence, we conclude that the enhanced X-ray flux from EX Lupi is a consequence of the increase in mass accretion rate, which was first detected in the optical outburst.

\section{Optical Variability During the 2008 Outburst}\label{optical_variability}

\begin{figure}
\centering
\begin{tabular}{cc}
\includegraphics[bb=220 -100 453 255,scale=0.55, angle=0]{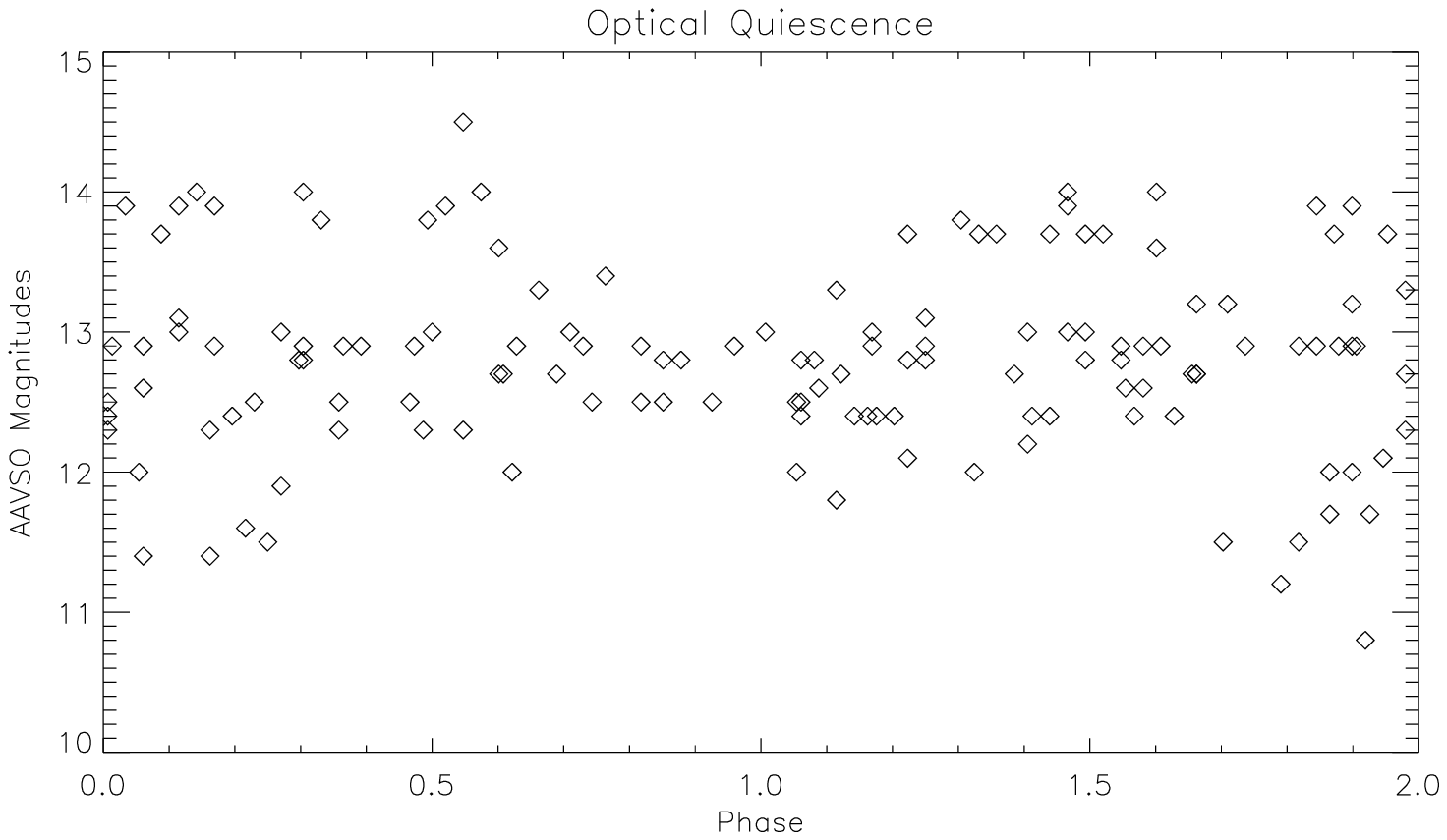} & \includegraphics[bb=380 -100 453 255,scale=0.55, angle=0]{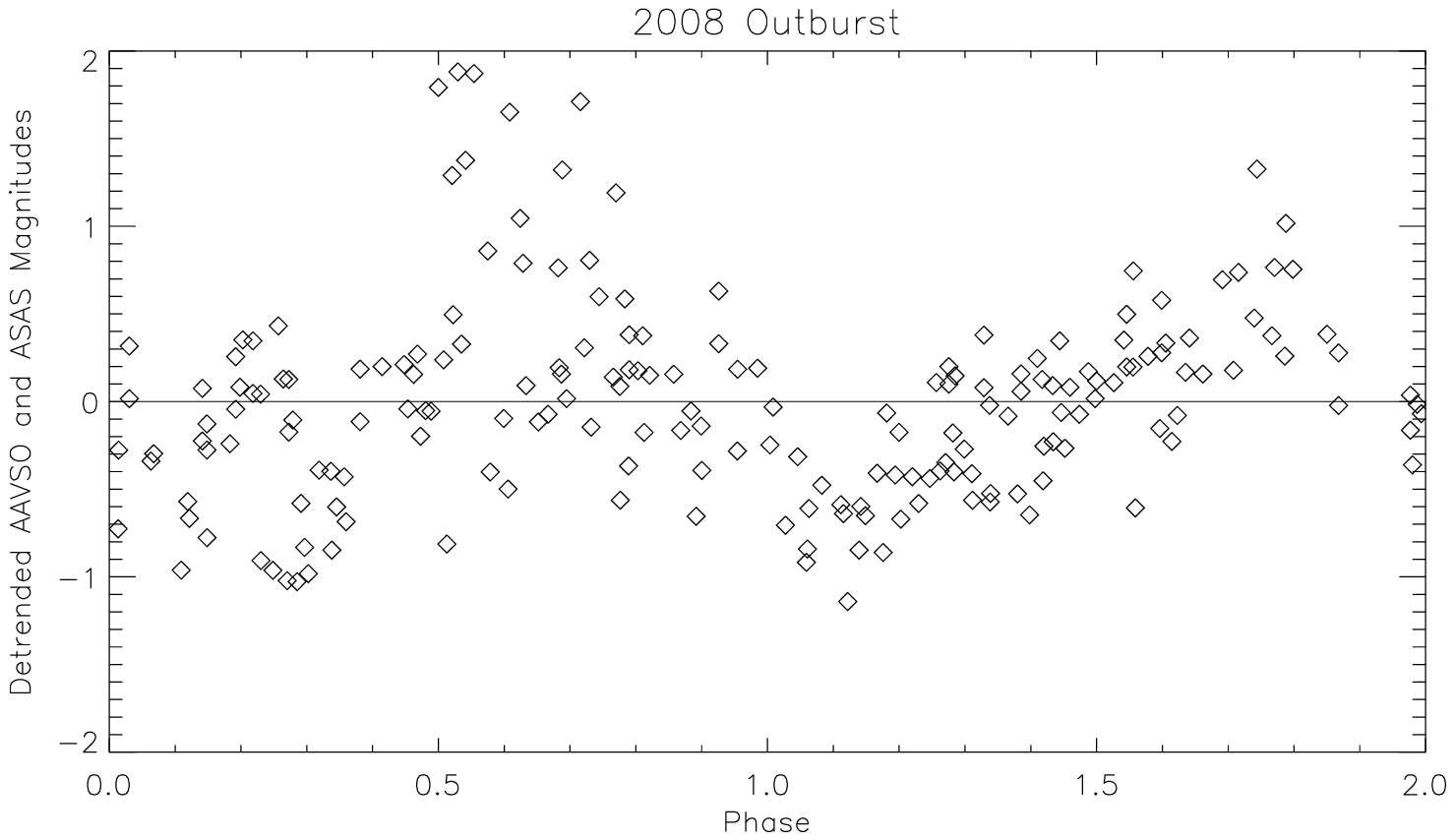} \\
\includegraphics[bb=30 -30 453 155,scale=.77, angle=0]{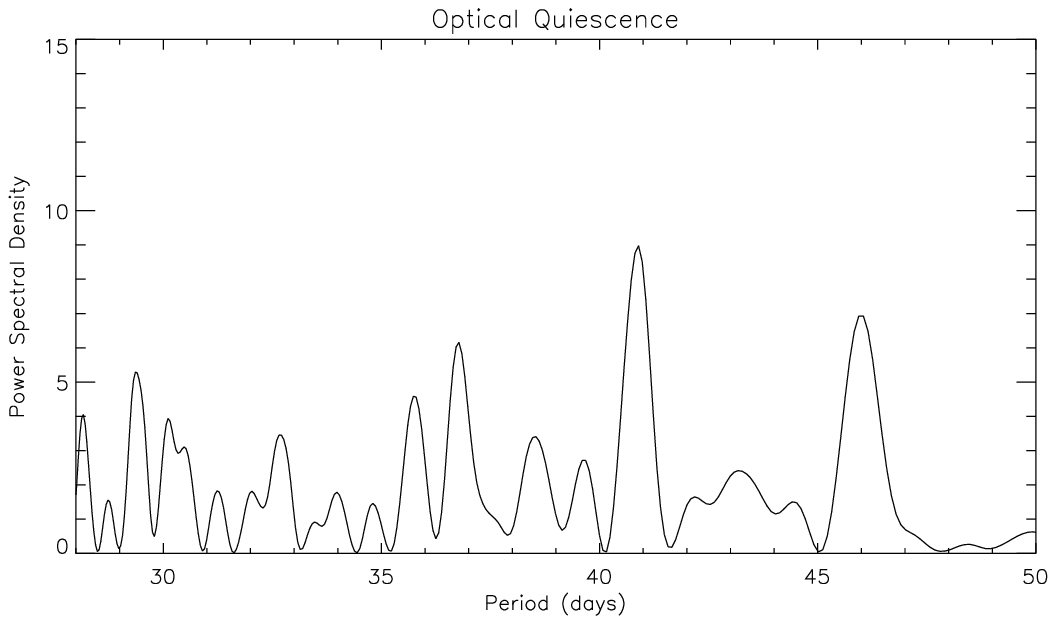} & \includegraphics[bb=140 -30 453 155,scale=0.77, angle=0]{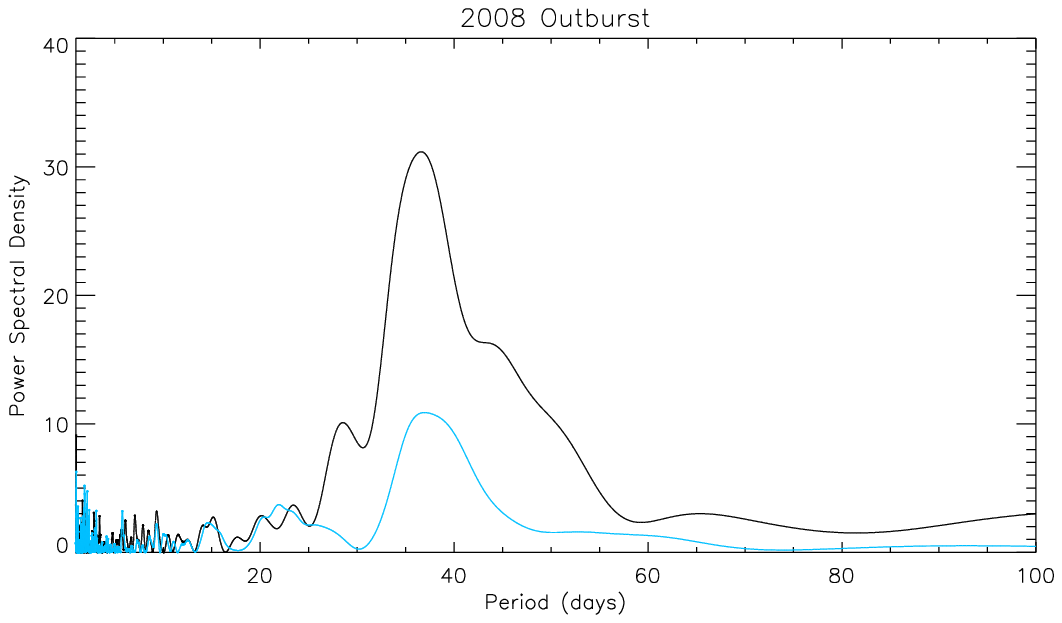} \\
\end{tabular}

\caption{Optical light curves of EX Lupi during quiescence before the 2008 optical outburst (top, left panel) and during the 2008 outburst (top, right panel) folded to two periods of 37 days.  The Lomb-Scargle periodogram of the quiescent optical light curve (bottom, left panel) shows no significant strong signals for periods between 28 and 50 days, while the Lomb-Scargle periodograms of the 2008 optical outburst light curve (ASAS and AAVSO data - black, ASAS data only - blue) shows a strong signal at a period of $\sim$37 days. 
The $\sim$37 day period is the only significant period in the outburst light curve between periods of 1 and 200 days and does not appear to be present during optical quiescence.
\label{folded_lcs_and_periodograms}}
\end{figure}

Figure \ref{crehr} shows V-band data from the All-Sky Automated Survey (ASAS) telescope and the American Association of Variable Star Observers (AAVSO) database.  ASAS data used in the figure were only those data flagged as photometric.  Close inspection of the EX Lupi optical light curve during the 2008 outburst reveals that EX Lupi appears to have exhibited a periodic variation in its optical flux, with a period of approximately 35 to 40 days.  We constructed Lomb-Scargle periodograms \citep{lom76,sca82,press89} to test this light curve for possible periodicities. 
With the ASAS magnitudes in our optical data set being of higher quality than the AAVSO magnitudes (since AAVSO data were obtained through visual estimations of the magnitude of EX Lupi), we first used only the ASAS data (48 data points from JD 2454504.8 to 2454697.6) in the periodicity analysis.  We subjected the data to the IDL routine \emph{linfit}, calculated a linear fit, and then ran the detrended data through an IDL Lomb-Scargle periodogram routine\footnote[3]{The IDL routine is available from Institut f\"{u}r Astronomie und Astrophysik at http://astro.uni-tuebingen.de/software/idl/aitlib/timing/scargle.html} to search through 20000 possible periods ranging from 1 to 200 days.  The results reveal a 37 day period that appears to be much more significant than all other possible periods identified in this analysis.  From the possible identified periods (Figure \ref{folded_lcs_and_periodograms}, bottom-right panel), the single strongest period appears at $\sim$37 days.  No other period has comparable strength. 
We reran the same analysis using both the ASAS and AAVSO data (the AAVSO data was selected to be within the same time range as that of the ASAS data).  In the resultant periodogram (Figure \ref{folded_lcs_and_periodograms}, bottom-right panel), the strongest signal, and in fact the only strong signal, corresponds to a period of $\sim$37 days. 
It therefore appears that the $\sim$37 day period, and only the $\sim$37 day period, is real. The top-right panel of Figure \ref{folded_lcs_and_periodograms} shows the 2008 optical outburst light curve of EX Lupi folded to a 37 day period.   Due to the increasing difference in optical magnitude between minima and maxima of the periodic light curve as the optical outburst progresses in 2008 January to August, the data points are subsequently spread in amplitude making the apparent 37 day periodicity difficult to observe in the folded light curve of the detrended data points.  However, there still appears to be a sinusoidal trend in the detrended data of the folded light curve --- as the folded light curve goes from phase 0.0 to phase 0.5 and back to phase 0.0, the majority of the detrended data lie below, then above, then below zero, respectively.

We note that in the above analysis, we searched for a periodicity in the optical light curve of EX Lupi only during the outburst phase, from 2008 January through October.  If this periodicity is present only during the outburst phase, it is probably associated in some way with accretion.  As a check for this, we applied the same analysis procedures to optical magnitudes of EX Lupi obtained from the AAVSO database over a period beginning about five years before the outburst and ending almost two years before the outburst (Julian dates 2451999.204 to 2453915). 

The Lomb-Scargle periodogram for the pre-outburst data shows no reliable peaks (Figure \ref{folded_lcs_and_periodograms}, bottom-left panel). 
The top-left panel of Figure \ref{folded_lcs_and_periodograms} shows the pre-outburst light curve data of EX Lupi folded to a period of 37 days.  Unlike the folded light curve of the detrended optical outburst data, which showed a small amplitude oscillation, we do not observe any sinusoidal trend to the folded light curve of the quiescent optical data.  Thus, the data show that the 37 day period that was present during the outburst, from January through October of 2008, was not detectable prior to the outburst. 

\begin{figure}
\centering
\includegraphics[scale=1.3, angle=0]{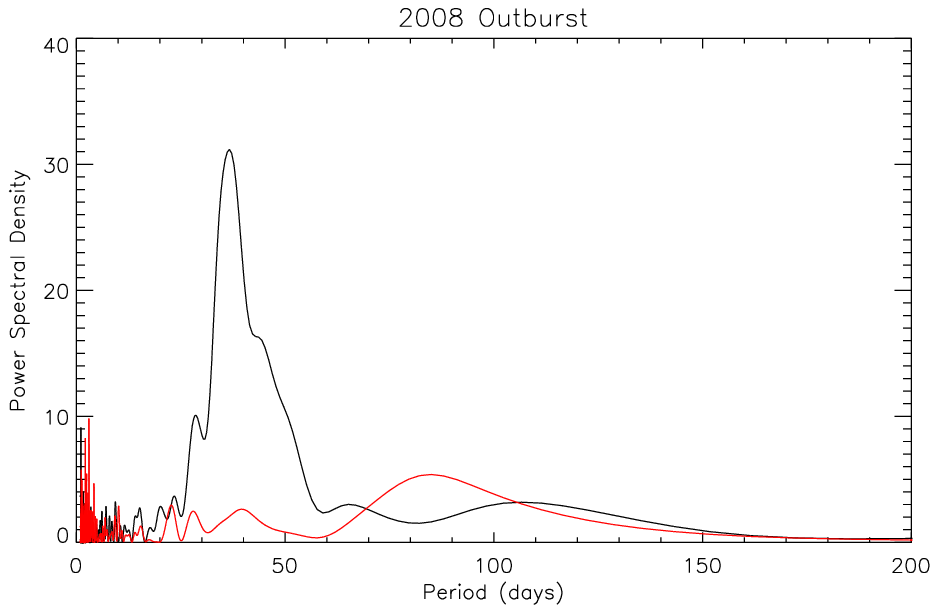}
\includegraphics[bb=-55 0 630 380,scale=.64, angle=0]{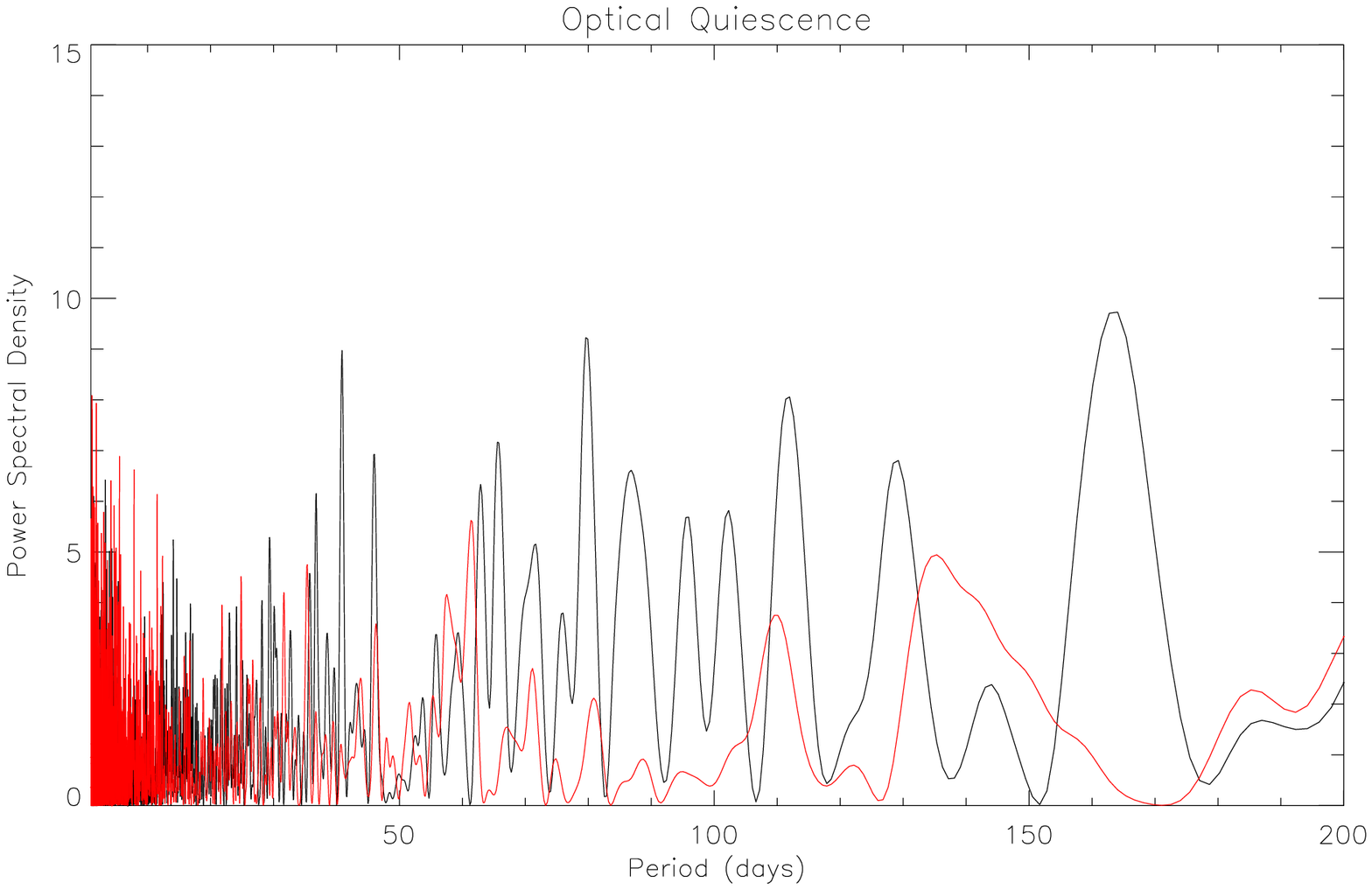}\\ 
\caption{EX Lupi optical light curve periodograms (black) overlaid with typical periodograms constructed from randomized light curve data (red) obtained during the 2008 outburst (top panel) and during quiescence before the 2008 optical outburst (bottom panel).  Outburst data periodograms use AAVSO and ASAS data while optical-quiescence periodograms use AAVSO data. 
The $\sim$37 day period observed during the 2008 optical outburst is apparently not randomly reproducable.\label{random_periodograms}} 
\end{figure}

As an additional check to the significance of the apparent $\sim$37 day periodicity, we randomized the aforementioned outburst and pre-outburst optical data and ran the same Lomb-Scargle periodogram analysis 1000 times per randomized dataset to determine if the 37 day period would show up in randomized datasets.  Figure \ref{random_periodograms} shows the original data periodograms (black) and a typical periodogram for the randomized datasets (red) during both the 2008 eruption (top panel) and during optical quiescence (bottom panel).  
None of the randomized light curves reproduce a reliable peak around the 37 day period (Figure \ref{random_periodograms}).  Therefore, the 37 day period cannot be randomly produced and is likely real in the optical outburst data.

\citet{dan12} interpreted the timescale of periodicity found in the light curve of this class of object (eruptive, rapidly-accreting pre-main sequence stars) in the context of mass possibly becoming ``trapped'' at the corotation radius of the accretion disk and cyclically accreting onto the star.  \citet{dan12} noted the $\sim$30 day periodicity during the outburst of EX Lupi, citing it as evidence for such ``mass trapping.''  Our further refines the likely periodicity timescale.

\section{Discussion \& Conclusions} 


The optical outbursts of eruptive objects such as EXors and FUors are thought to be the result of the sudden onset of large-scale accretion events.  Accretion of circumstellar gas should result in the production of a relatively low-temperature, X-ray-generating plasma as free-falling material impacts the stellar photosphere and is shock-heated to temperatures of a few million Kelvin at the footprint of the accretion stream footprint in the stellar photosphere.  If the intervening hydrogen column density is low enough and the emission measure of the shock-heated plasma is large enough, then we should be able to detect this X-ray flux component  -- as is clearly the case for EX Lupi -- as well as the hotter coronal X-ray emission resulting from non-accretion-related magnetic activity (see \citealt{gro10}).  As the accretion subsides and the X-ray output of the accretion footprint diminishes, the softer X-ray flux should decrease, resulting in the median photon energy increasing, the hardness ratio becoming less negative, and an overall decrease in X-ray flux.  Given that all three of these correlations are observed in the X-ray spectrum of EX Lupi as time progresses and that the optical flux levels of EX Lupi return to quiescent levels as the accretion event subsides, the hypothesis that the elevated X-ray flux observed during the 2008 outburst was generated via circumstellar accretion onto the stellar photosphere is very strongly supported.

Similar correlations were documented in optical, near-infrared, and X-ray observations of the young, PMS star V1647 Ori when it was observed to erupt in 2003 and then again in 2008 \citep{tee11}.  In the case of V1647 Ori, however, the CXO spectra showed no evidence for the soft X-ray component that should be generated by accretion. \citet{tee11} interpreted the large increase and subsequent decrease in hard X-ray flux from V1647 Ori to be the result of accretion and explained the absence of an observed soft X-ray component during the accretion episode as a consequence of very high intervening hydrogen column density.  The observed increase in the hard X-ray flux was thought to originate in the accretion funnels from magnetic reconnection events.  As accretion subsided, the hard X-ray flux decreased as well.  

During the 2008 outburst of EX Lupi, we first observed a two-component plasma, with one component consisting of a plasma with a temperature characteristic of shock-heated accreting material and the other component consisting of a plasma with a temperature much higher and characteristic of those temperatures found in the V1647 Ori observations by CXO \citep{kas04,kas06,tee11}, XMM-Newton \citep{gro05}, and Suzaku \citep{ham10}.  As time passed and the accretion subsided, the X-ray spectrum of EX Lupi changed such that the lower-temperature component emission measure and flux level decreased while the hotter plasma component's emission measure and flux level generally decreased as well.  After the accretion episode ended, the spectrum was characterized by faint emission from hotter plasma characteristic of coronal activity.  This sequence of changes follows the expected pattern of spectral changes if the elevated X-ray flux is accretion shock-generated and the column density is low enough that we can detect soft X-rays generated in or near the photosphere. In addition, an elevated level of harder X-rays during the optical outburst may also be associated with magnetic reconnection events in or near the accretion-fed loops and/or 
the accretion funnels.

\clearpage

\newpage
\begin{deluxetable}{llcccccc}
\tabletypesize{\scriptsize}
\tablewidth{0pt}
\tablecaption{\label{tab-1}Chandra ACIS Observations of EX Lupi Following the 2008 January Outburst}
\tablehead{\colhead{ObsID} & \colhead{Observation} & \colhead{JD} & \colhead{Exposure} & \colhead{Net} & \colhead{Mean Count} & \colhead{Median Photon}  & \colhead{Hardness Ratio\tablenotemark{a}} \\
\colhead{ } & \colhead{Date} & \colhead{ } & \colhead{(ks)} & \colhead{Counts} & \colhead{Rate (cts ks$^{-1}$)} & \colhead{Energy (keV)}  & \colhead{(H--S/H+S)} }
\startdata

8923 & 2008 Mar 25 & 2454551& 20.1 & 944 & 47.6 $\pm$ 1.6 & 1.15 $\pm$ 0.02 & $-$0.06 $\pm$ 0.03\\
8924 & 2008 Jun 16 & 2454634 & 20.1 & 318 & 16.0 $\pm$ 0.9 & 1.09 $\pm$ 0.03 &  $-$0.13 $\pm$ 0.06\\
8925\tablenotemark{b} & 2008 Oct 6 & 2454746 & 30.1 & 232 & 7.7 $\pm$ 0.5 & 1.69 $\pm$ 0.06 &  0.55 $\pm$ 0.05\\
10789\tablenotemark{b} & 2008 Oct 6 & 2454746\\
10791\tablenotemark{b} & 2008 Oct 9 & 2454749\\

\enddata

\tablenotetext{a}{Hardness ratio computed using the total numbers of hard and soft X-ray photons from the entire observation.}
\tablenotetext{b}{Parameter values are derived from the combined spectra of the three 2008 October observations (ObsIDs 8925, 10789, and 10791.}
\tablecomments{All errors are 1$\sigma$.  The net counts for each observation are the total number of counts within the 0.3--10.0 keV range.  Median photon energy uncertainties were calculated via the half-sample method used in \citet{kas06}.  Mean count rates were determined by dividing the net counts by exposure times.  Uncertainties for mean count rates and hardness ratios of total counts follow Poisson statistics. }
\end{deluxetable}

\newpage
\begin{deluxetable}{lllcccccccc}
\rotate
\setlength{\tabcolsep}{0.02in} 
\tabletypesize{\scriptsize}
\tablewidth{0pt}
\tablecaption{\label{tab-3}Best-Fit Models for EX Lupi Observations}
\tablehead{\colhead{ObsID} & \colhead{Observation} & \colhead{Observation} & \colhead{Reduced} & \colhead{Degrees of }  & \colhead{Plasma} & \colhead{N$_{H}$} & \colhead{kT$_{X}$} & \colhead{EM$_{}$}& \colhead{Plasma Component F$_{X}$} & \colhead{Total Observed L$_{X}$}  \\ 
\colhead{ } & \colhead{Date} & \colhead{Time} & \colhead{$\chi$$^{2}$ } & \colhead{Freedom } & \colhead{Component} & \colhead{($\times$ 10$^{22}$ cm$^{-2}$)} & \colhead{(keV)} & \colhead{($\times$ 10$^{53}$ cm$^{-3}$)} & \colhead{ ($\times$ 10$^{-13}$ ergs cm$^{-2}$ s$^{-1}$)} & \colhead{($\times$ 10$^{29}$ ergs s$^{-1}$) } }

\startdata 
8923 & 2008 Mar 25 & 22:29:38 & 0.92 & 90 & 1 & 0.4${+0.2\atop-0.2}$ & 0.4${+0.3\atop-0.1}$  & 3.1${+4.9\atop-2.0}$                               		  & 	9.8			& 		8.5		\\
& & & & & 2 & \nodata & 1.7${+0.6\atop-0.4}$  & 0.7${+0.4\atop-0.2}$                												  & 	2.5			&				\\

\\

8924 & 2008 Jun 16 & 22:06:06& 0.90 & 42 & 1& 0.01${+0.13\atop-0.01}$ & 0.5${+0.1\atop-0.1}$ & 0.2${+0.1\atop-0.1}$            		 		 & 	0.8			& 		4.6		\\                 
&& & &  					     & 2 & 1.7${+3.5\atop-0.8}$ & 0.7${+1.0\atop-0.4}$ & 0.9${+17.6\atop-0.7}$ 	 					 & 	3.2			&				\\
& && &  					     & 3 & $\sim$40 & $\sim$4 & $\sim$1 	 													& 	5.7			&				\\
\\
XMM\tablenotemark{a} & 2008 Aug 10 && 0.87 & 12 & 1 & 0.04${+0.11\atop-0.04}$ & 		0.4${+0.3\atop-0.1}$  & 0.04${+0.04\atop-0.01}$ 			& 	0.1			&		1.6		\\
             
& & &&   &    2 &                                        2.7${+4.6\atop-2.3}$ & 		4.6${+N/A\atop-3.0}$  & 0.2${+1.0\atop-0.1}$ 		     		& 	1.1			&				\\
        
\\

8925\tablenotemark{b}  & 2008 Oct 6&00:01:53 & 0.92 & 35 & 1&	 0.9${+0.4\atop-0.4}$ & 		0.9${+0.3\atop-0.4}$  & 0.5${+1.0\atop-0.4}$ 				& 	1.8			&		1.9		\\
  
10789\tablenotemark{b}  & 2008 Oct 6&17:32:36&& & 				2&		3.8${+13.5\atop-3.0}$ & 		2.3${+27.1\atop-1.4}$  & 0.3${+1.1\atop-0.3}$  				& 	1.4			&	\\			
10791\tablenotemark{b}  & 2008 Oct 9&12:08:54\\

%
%

 \enddata
\tablenotetext{a}{Here we only report the spectral model of the low-level period (first row of Table 1 of \citet{gro10}).}
\tablenotetext{b}{Parameter values are derived from the combined spectra of the three 2008 October observations.}
\tablecomments{Uncertainties given for hydrogen column density and plasma temperature correspond to the 90$\%$ confidence intervals.  Emission measures and luminosities assume a distance of 155 pc to EX Lupi. X-ray fluxes for each of the plasma components (tenth column) have been corrected for absorption; however, the total observed X-ray luminosities are not corrected for absorption and give the total X-ray flux derived from the modeling procedure.  X-ray fluxes and luminosities are derived for the 0.2--10.0 keV energy range.  All chemical abundances are fixed at XEST values. Due to poor constraints on the third plasma component of the 2008 June spectrum model, the error ranges have been omitted and the best-fit values are given only.  }

\end{deluxetable}

\newpage


\begin{rotate}
\begin{figure}

\centering
\includegraphics[scale=0.7, bb=160 0 822 332]{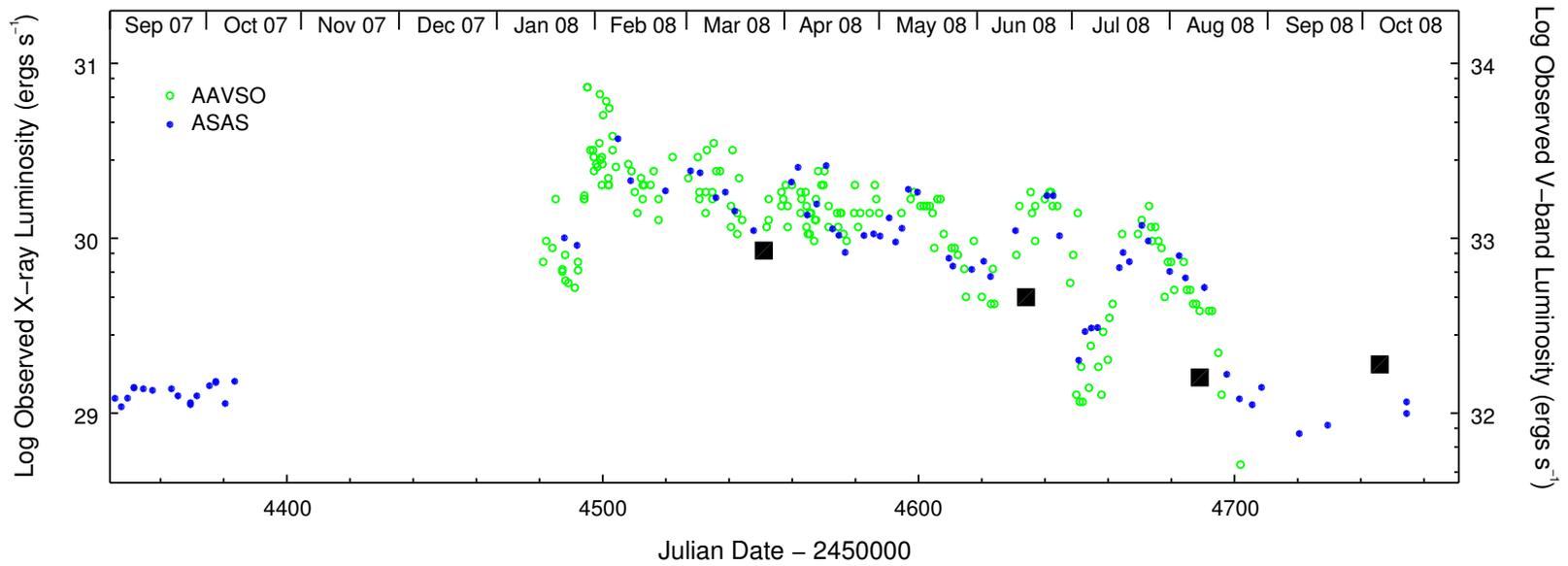}\ 
\caption{Optical and X-ray light curves of EX Lupi during the 2008 outburst.  After the onset of the optical outburst of EX Lupi, the observed X-ray (crosses) luminosity decreased as the V-band (blue and green circles) luminosity decreased.  Error bars (one sigma) are typically smaller than the plotted symbols.  During the 2008 outburst, the optical light curve shows strong evidence for a $\sim$37 day periodicity.}
\label{crehr}
\end{figure}
\end{rotate}

\end{document}